\documentclass[twocolumn,showpacs,preprintnumbers,amsmath,amssymb]{revtex4}

\usepackage{graphicx}% Include figure files
\usepackage{dcolumn}% Align table columns on decimal point
\usepackage{bm}% bold math
\usepackage{epsf,epsfig,subfigure,axodraw,graphicx,color}

\begin{document}

\preprint{SNUTP 07-012}

\title{Axion as a CDM component}% Force line breaks with \\
\author{Jihn E. Kim}

% \email{Second.Author@institution.edu}
\affiliation{%
Department of Physics and Astronomy and Center for Theoretical Physics,\\
Seoul National University, Seoul 151-747, Korea}

\begin{abstract}
I discuss the essential features of the QCD axion: the strong CP solution and hence its theoretical necessity. I also review the effects of the QCD axion on astrophysics and cosmology, in particular with emphasis on its role in the dark matter component together with its supersymmetric partner axino. It is pointed out that string theory may or may not give a detectable QCD axion.
\end{abstract}



\maketitle

\section{ Introduction}
Cosmology with cold dark matter (CDM) was the leading candidate of the early universe in 1980s \cite{Blum}. But this view has given a way to the new cosmology with the discovery of dark
energy (DE) in 1998 \cite{DE98}. The current view of the dominant components of the universe is $\Omega_{\rm CDM}\simeq 0.23$ and $\Omega_\Lambda\simeq 0.73$ \cite{WMAP3}. The plausible DM candidates at present are the lightest SUSY particle (LSP), axion, axino, and gravitino. Here, we will review mostly on axion and its CDM-related possibility. The need for DM was first suggested by Zwicky in 1933 \cite{Zwicky}.

There are numerous cases supporting the nonluminous dark matter in the universe: flat rotation curves, Chandra satellite photo, and gravitational lensing effects. If the galactic bulge is the dominant mass in the galaxy, the velocity $v$ of a star located at $r$ from the center behaves as $v\sim r^{-1/2}$. But the observed flat rotation curve \cite{flatrot} violates this expectation and implies the extended mass into the halo as $\rho(r)\sim 1/r^2$. Also, the Chandra observation of X-ray and gravitational lensing images implies the matter profile around the bullet galaxy \cite{Chandra}. The circular gravitational lensing photo \cite{gravlens} also supports the idea that DM exist. The DM density around us here is usually taken as
$\rho_{\rm DM}\simeq 0.3$ GeV/cm$^3$.

\begin{figure}
\begin{center}
\begin{picture}(400,260)(0,-30)
\Line(30,-20)(30,219.1) \Line(230,-20)(230,219.1)
\Line(30,-20)(230,-20) \Line(30,219.1)(230,219.1)
\Text(130,-40)[c]{Log($m_\nu$[GeV])}
\rText(6,120)[t][l]{Log($\Omega_{\nu}h^2$)}

\Text(27,0)[r]{$-1$} \Text(27,40)[r]{$0$} \Text(27,80)[r]{$1$}
\Text(27,120)[r]{$2$} \Text(27,160)[r]{$3$}\Text(27,200)[r]{$4$}
 \Line(30,212)(33,212) \Line(230,212)(227,212)
 \Line(30,200)(36,200)  \Line(230,200)(224,200)
 \Line(30,188)(35,188) \Line(230,188)(225,188)
 \Line(30,184.1)(33,184.1) \Line(230,184.1)(227,184.1)
 \Line(30,179.1)(33,179.1) \Line(230,179.1)(227,179.1)
 \Line(30,172)(33,172) \Line(230,172)(227,172)
 \Line(30,160)(36,160)  \Line(230,160)(224,160)
 \Line(30,148)(35,148) \Line(230,148)(225,148)
 \Line(30,144.1)(33,144.1) \Line(230,144.1)(227,144.1)
 \Line(30,139.1)(33,139.1) \Line(230,139.1)(227,139.1)
 \Line(30,132)(33,132) \Line(230,132)(227,132)
 \Line(30,120)(36,120)  \Line(230,120)(224,120)
 \Line(30,108)(35,108) \Line(230,108)(225,108)
 \Line(30,104.1)(33,104.1) \Line(230,104.1)(227,104.1)
 \Line(30,99.1)(33,99.1) \Line(230,99.1)(227,99.1)
 \Line(30,92)(33,92) \Line(230,92)(227,92)
 \Line(30,80)(36,80)  \Line(230,80)(224,80)
 \Line(30,68)(35,68) \Line(230,68)(225,68)
 \Line(30,64.1)(33,64.1) \Line(230,64.1)(227,64.1)
 \Line(30,59.1)(33,59.1) \Line(230,59.1)(227,59.1)
 \Line(30,52)(33,52) \Line(230,52)(227,52)
 \Line(30,40)(36,40)  \Line(230,40)(224,40)
 \Line(30,28)(35,28) \Line(230,28)(225,28)
 \Line(30,24.1)(33,24.1) \Line(230,24.1)(227,24.1)
 \Line(30,19.1)(33,19.1) \Line(230,19.1)(227,19.1)
 \Line(30,12)(33,12) \Line(230,12)(227,12)
 \Line(30,0)(36,0)  \Line(230,0)(224,0)
 \Line(30,-12)(35,-12) \Line(230,-12)(225,-12)
 \Line(30,-15.9)(33,-15.9) \Line(230,-15.9)(227,-15.9)

\Text(50,-27)[c]{${-7}$}\Text(70,-27)[c]{${-6}$}
\Text(90,-27)[c]{${-5}$}\Text(110,-27)[c]{${-4}$}
\Text(130,-27)[c]{${-3}$}\Text(150,-27)[c]{${-2}$}
\Text(170,-27)[c]{${-1}$}\Text(190,-27)[c]{$0$}
\Text(210,-27)[c]{${1}$}

 \Line(44,-20)(44,-17) \Line(44,219.1)(44,216.1)
 {\SetWidth{1}\SetColor{Blue} \Line(50,-20)(50,-14)
 \Line(50,219.1)(50,213.1) \Text(50,-9)[c]{$10^2$eV}}
 \Line(64,-20)(64,-17) \Line(64,219.1)(64,216.1)
 \Line(70,-20)(70,-14)  \Line(70,219.1)(70,213.1)
 \Line(84,-20)(84,-17) \Line(84,219.1)(84,216.1)
 \Line(90,-20)(90,-14)  \Line(90,219.1)(90,213.1)
 \Line(104,-20)(104,-17) \Line(104,219.1)(104,216.1)
 \Line(110,-20)(110,-14)  \Line(110,219.1)(110,213.1)
 \Line(124,-20)(124,-17) \Line(124,219.1)(124,216.1)
 \Line(130,-20)(130,-14)  \Line(130,219.1)(130,213.1)
 \Line(144,-20)(144,-17) \Line(144,219.1)(144,216.1)
 \Line(150,-20)(150,-14)  \Line(150,219.1)(150,213.1)
 \Line(164,-20)(164,-17) \Line(164,219.1)(164,216.1)
 \Line(170,-20)(170,-14)  \Line(170,219.1)(170,213.1) \Line(184,-20)(184,-17) \Line(184,219.1)(184,216.1)
 {\SetWidth{1}\SetColor{Blue} \Line(190,-20)(190,-14)
 \Line(190,219.1)(190,213.1) \Text(190,-9)[c]{GeV}}
 \Line(204,-20)(204,-17) \Line(204,219.1)(204,216.1)
 \Line(210,-20)(210,-14)  \Line(210,219.1)(210,213.1)
 \Line(224,-20)(224,-17) \Line(224,219.1)(224,216.1)

 %Curves
 {\SetWidth{1} \Line(30,2)(125.6,191.1)
 \Curve{(125.6,191.1)(136,196.8)(145.6,191.1)}
  {\SetColor{Blue}
 \Curve{(145.6,191.1)(157.6,164)(172.3,120)
 (206,0)(211,-20)}}%Dirac
 {\SetColor{Red}
 \Curve{(157.2,191.1)(178,124)(202,42)(213,0)
 (217.5,-20)}%Majorana
 \Curve{(125.6,191.1)(139.8,203.9)(157.1,191.1)}
 }
 {\SetColor{Brown}
 \DashLine(30,2)(230,2){4}
 }
 }
 \Text(185,70)[r]{Dirac}
 \Text(176,140)[l]{Majorana}

\end{picture}
\end{center}
\caption{A schematic view of the Lee-Weinberg calculation of $\Omega_\nu h^2$. For the larger mass region, the freezeout temperature is $T_f\approx m_\nu/20$ [8]. The dash line is for $\Omega_\nu h^2=0.113$ [9].}\label{LeeWein}
\end{figure}
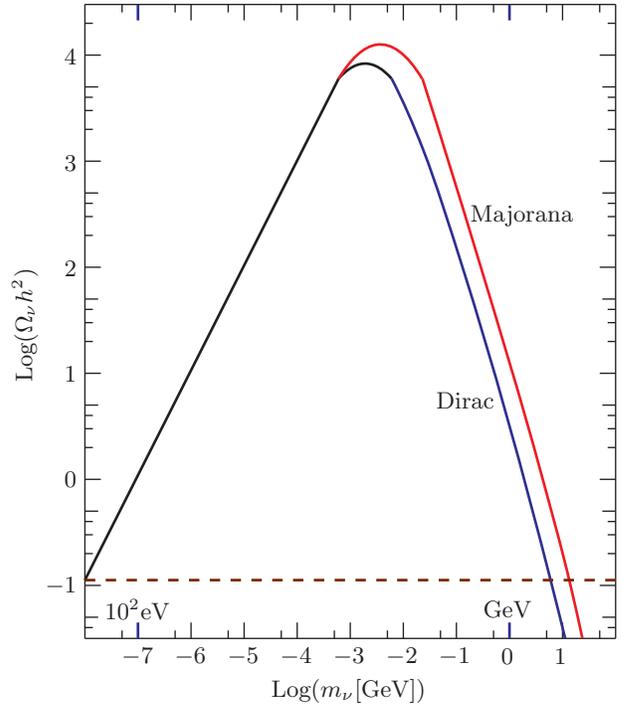

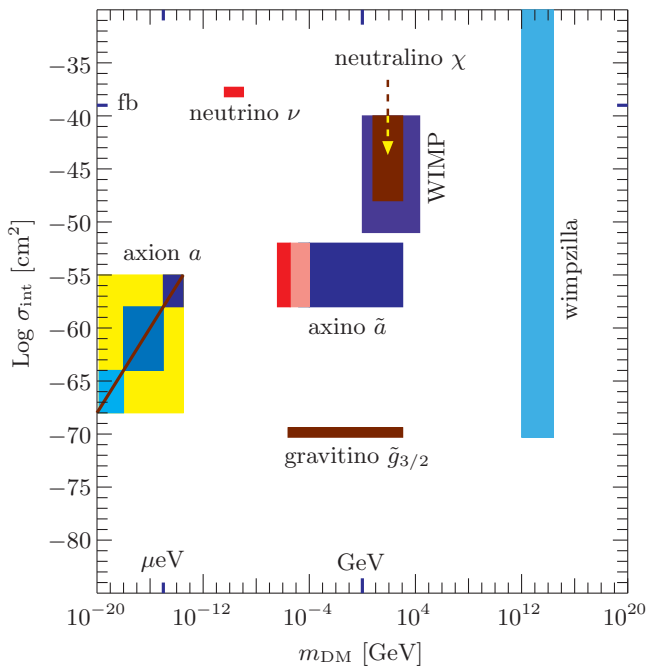
\begin{figure}
\begin{center}
\begin{picture}(400,240)(10,-10)
\SetScale{0.8}
\Line(50,10)(300,10) \Line(50,10)(50,285)
 \rText(10,150)[t][l]{Log~$\sigma_{\rm int}$ [cm$^2$]}
 \Text(140,-15)[c]{$m_{\rm DM}$ [GeV]}
\Line(300,10)(300,285)\Line(50,285)(300,285)

\Line(50,280)(55,280)\Line(295,280)(300,280)
\Line(50,275)(55,275)\Line(295,275)(300,275)
\Line(50,270)(55,270)\Line(295,270)(300,270)
\Line(50,265)(55,265)\Line(295,265)(300,265)
\Line(50,260)(57,260)\Text(38,207.5)[r]{${-35}$}
\Line(293,260)(300,260)
      \CBox(110,248.5)(119,244){Red}{Red}
\Line(50,255)(55,255)\Line(295,255)(300,255)
\Line(50,250)(55,250)\Line(295,250)(300,250)
\Line(50,245)(55,245)\Line(295,245)(300,245)
{\SetWidth{1.5}\SetColor{Blue}
 \Line(50,240)(55,240)\Line(295,240)(300,240)}
\Text(48,194)[l]{fb}
\Line(50,235)(57,235)\Text(38,187.5)[r]{${-40}$}
\Line(293,235)(300,235)
\Line(50,230)(55,230)\Line(295,230)(300,230)
\Line(50,225)(55,225)\Line(295,225)(300,225)
     \CBox(175,235)(202,180){BlueViolet}{BlueViolet}
     \CBox(180,235)(194,195){Brown}{Brown}
\Line(50,220)(55,220)\Line(295,220)(300,220)
\Line(50,215)(55,215)\Line(295,215)(300,215)
\Line(50,210)(57,210)\Text(38,167.5)[r]{${-45}$}
\Line(293,210)(300,210)
\Line(50,205)(55,205)\Line(295,205)(300,205)
\Line(50,200)(55,200)\Line(295,200)(300,200)
\Line(50,195)(55,195)\Line(295,195)(300,195)
\Line(50,190)(55,190)\Line(295,190)(300,190)
\Line(50,185)(57,185)\Text(38,147.5)[r]{${-50}$}
\Line(293,185)(300,185)
\Line(50,180)(55,180)\Line(295,180)(300,180)
\Line(50,175)(55,175)\Line(295,175)(300,175)
    \CBox(51,95)(90.6,160){Yellow}{Yellow}
   \CBox(81.25,145)(90.6,160){Blue}{Blue}
    \CBox(62.5,115)(81.25,145){RoyalBlue}{RoyalBlue}
    \CBox(51,95)(62.5,115){Cyan}{Cyan}
    {\SetColor{Brown}\SetWidth{1.5}\Line(50,95)(90.6,160)}
   \CBox(150,175)(194,145){Blue}{Blue}
   \CBox(145,175)(150,145){RoyalBlue}{RoyalBlue}
  \CBox(141,175)(150,145){Salmon}{Salmon}
    \CBox(135,175)(141,145){Red}{Red}
\Line(50,170)(55,170)\Line(295,170)(300,170)
\Line(50,165)(55,165)\Line(295,165)(300,165)
\Line(50,160)(57,160)\Text(38,127.5)[r]{${-55}$}
\Line(293,160)(300,160)
\Line(50,155)(55,155)\Line(295,155)(300,155)
\Line(50,150)(55,150)\Line(295,150)(300,150)
\Line(50,145)(55,145)\Line(295,145)(300,145)
\Line(50,140)(55,140)\Line(295,140)(300,140)
\Line(50,135)(57,135)\Text(38,107.5)[r]{${-60}$}
\Line(293,135)(300,135)
\Line(50,130)(55,130)\Line(295,130)(300,130)
     \CBox(140,88)(194,83.5){Brown}{Brown}
     \CBox(250,285)(265,83.5){CornflowerBlue}{CornflowerBlue}
\Line(50,125)(55,125)\Line(295,125)(300,125)
\Line(50,120)(55,120)\Line(295,120)(300,120)
\Line(50,115)(55,115)\Line(295,115)(300,115)
\Line(50,110)(57,110)\Text(38,87.5)[r]{${-65}$}
\Line(293,110)(300,110)
\Line(50,105)(55,105)\Line(295,105)(300,105)
\Line(50,100)(55,100)\Line(295,100)(300,100)
\Line(50,95)(55,95)\Line(295,95)(300,95)
\Line(50,90)(55,90)\Line(295,90)(300,90)
\Line(50,85)(57,85)\Text(38,67.5)[r]{${-70}$}
\Line(293,85)(300,85)
\Line(50,80)(55,80)\Line(295,80)(300,80)
\Line(50,75)(55,75)\Line(295,75)(300,75)
\Line(50,70)(55,70)\Line(295,70)(300,70)
\Line(50,65)(55,65)\Line(295,65)(300,65)
\Line(50,60)(57,60)\Text(38,47.5)[r]{${-75}$}
\Line(293,60)(300,60)
\Line(50,55)(55,55)\Line(295,55)(300,55)
\Line(50,50)(55,50)\Line(295,50)(300,50)
\Line(50,45)(55,45)\Line(295,45)(300,45)
\Line(50,40)(55,40)\Line(295,40)(300,40)
\Line(50,35)(57,35)\Text(38,27.5)[r]{ ${-80}$}\Line(293,35)(300,35)
\Line(50,30)(55,30)\Line(295,30)(300,30)
\Line(50,25)(55,25)\Line(295,25)(300,25)
\Line(50,20)(55,20)\Line(295,20)(300,20)
\Line(50,15)(55,15)\Line(295,15)(300,15)

 \Text(40,0)[c]{$10^{-20}$}
\Line(56.25,10)(56.25,15)\Line(56.25,285)(56.25,280)
\Line(62.5,10)(62.5,15)\Line(62.5,285)(62.5,280)
\Line(68.75,10)(68.75,15)\Line(68.75,285)(68.75,280)
\Line(75,10)(75,17)\Line(75,285)(75,278)
{\SetColor{Blue}\SetWidth{1.5}\Line(81.25,10)(81.25,15)
\Line(81.25,285)(81.25,280)}
\Text(65,20)[c]{$\mu$eV}
\Line(87.5,10)(87.5,15)\Line(87.5,285)(87.5,280)
\Line(93.75,10)(93.75,15)\Line(93.75,285)(93.75,280)
\Line(100,10)(100,19)\Line(100,285)(100,276)
\Text(80,0)[c]{$10^{-12}$}
\Line(106.25,10)(106.25,15)\Line(106.25,285)(106.25,280)
\Line(112.5,10)(112.5,15)\Line(112.5,285)(112.5,280)
\Line(118.75,10)(118.75,15)\Line(118.75,285)(118.75,280)
\Line(125,10)(125,17)\Line(125,285)(125,278)
\Line(131.25,10)(131.25,15)\Line(131.25,285)(131.25,280)
\Line(137.5,10)(137.5,15)\Line(137.5,285)(137.5,280)
\Line(143.75,10)(143.75,15)\Line(143.75,285)(143.75,280)
\Text(120,0)[c]{$10^{-4}$}\Line(150,10)(150,19)
\Line(150,285)(150,276)
\Line(156.25,10)(156.25,15)\Line(156.25,285)(156.25,280)
\Line(162.5,10)(162.5,15)\Line(162.5,285)(162.5,280)
\Line(168.75,10)(168.75,15)\Line(168.75,285)(168.75,280)
{\SetColor{Blue}\SetWidth{1.5}\Line(175,10)(175,17)
\Line(175,285)(175,278)}
\Text(140,20)[c]{GeV}
\Line(181.25,10)(181.25,15)\Line(181.25,285)(181.25,280)
\Line(187.5,10)(187.5,15)\Line(187.5,285)(187.5,280)
\Line(193.75,10)(193.75,15)\Line(193.75,285)(193.75,280)
\Line(200,10)(200,19)\Line(200,285)(200,276)
\Text(160,0)[c]{$10^{4}$}
\Line(206.25,10)(206.25,15)\Line(206.25,285)(206.25,280)
\Line(212.5,10)(212.5,15)\Line(212.5,285)(212.5,280)
\Line(218.75,10)(218.75,15)\Line(218.75,285)(218.75,280)
\Line(225,10)(225,17)\Line(225,285)(225,278)
\Line(231.25,10)(231.25,15)\Line(231.25,285)(231.25,280)
\Line(237.5,10)(237.5,15)\Line(237.5,285)(237.5,280)
\Line(243.75,10)(243.75,15)\Line(243.75,285)(243.75,280)
\Text(200,0)[c]{$10^{12}$}\Line(250,10)(250,19)
\Line(250,285)(250,276)
\Line(256.25,10)(256.25,15)\Line(256.25,285)(256.25,280)
\Line(262.5,10)(262.5,15)\Line(262.5,285)(262.5,280)
\Line(268.75,10)(268.75,15)\Line(268.75,285)(268.75,280)
\Line(275,10)(275,17)\Line(275,285)(275,278)
\Line(281.25,10)(281.25,15)\Line(281.25,285)(281.25,280)
\Line(287.5,10)(287.5,15)\Line(287.5,285)(287.5,280)
\Line(293.75,10)(293.75,15)\Line(293.75,285)(293.75,280)
\Line(300,10)(300,17)\Line(300,285)(300,253)
\Text(240,0)[c]{$10^{20}$}

%%%%%%%%Particles

 \SetWidth{0.5}
  \Text(75,190)[l]{neutrino $\nu$}
 \Text(50,137)[l]{axion $a$}
 \Text(120,109)[l]{axino ${\tilde a}$}
 \Text(111,59)[l]{gravitino ${\tilde g_{3/2}}$}
 \rText(167,183)[t][l]{WIMP}
  \Text(180,210)[r]{neutralino ${\chi}$}
 \rText(218,150)[t][l]{wimpzilla}
 {\SetWidth{1.2}\SetColor{Brown}\DashLine(187,252)(187,235){3}
 \SetColor{Yellow}\DashLine(187,235)(187,219){3}
 \LongArrow(187,222)(187,218)}

\end{picture}
\end{center}
\caption{Some proposed particles in the interaction cross section versus mass
plane.}\label{sigmavsmass}
\end{figure}

The proposed CDM candidates belong to either incoherent particles or coherent oscillations of spin-0 fields. The incoherent CDM particles are usually called weakly interacting massive particles(WIMP) or decay products of WIMPs. A coherent oscillation arises from the potential describing a bosonic particle. WIMPs are the massive particles with weak interaction cross sections, which was first discussed by B. W. Lee and S. Weinberg  in 1977 in terms of a heavy neutrino as shown in Fig. \ref{LeeWein} \cite{LeeWein}.
This implies that the usage $\lq\lq$weak"
is involved in WIMP. SUSY with the R-parity conservation allows the LSP as such a particle. The LSP interaction is $\lq\lq$weak" since the interaction mediators (SUSY particles) are supposed to be in the 100 GeV range. For a WIMP to be a successful CDM candidate, usually the WIMP interaction cross section at the time of decoupling is needed around \cite{WMAPsigma}
\begin{align}
&\langle\sigma_{\rm int} v\rangle|_{\rm at\ decoupling}\approx 0.2\times
10^{-26} ~{\rm cm^3s}^{-1},\nonumber\\
&\quad\quad {\rm with}\ \Omega_m h^2\simeq 0.113\pm 0.009.\label{WIMPcond}
\end{align}
This is roughly explained by the LSP with the low energy SUSY, which is the reason that we are so much concerned about the WIMP LSP. The proposed particles are shown in the $\sigma_{\rm int}$ versus mass plane in Fig. \ref{sigmavsmass} with minor modification from that of \cite{Roszfig}. Currently, there are experimental efforts to discover the LSP as predicted
by SUSY models. Also, direct cosmological searches are going on
\cite{Dama}. At the LHC, the probable LSP mass ranges will be looked for.
%%%%%%%%%%%%%%%%%%%%%%%%%%%%%%%%%%%%%%%%%%%%%%%%%%%%%%%%%%%%%%%%%%%%%%%%%%%%%%
%%%%%%%%%%%%%%%%%%%%%%%%%%%%%%%%%%%%%%%%%%%%%%%%%%%%%%%%%%%%%%%%%%%%%%%%%%%%%%

%%%%%%%%%%%%%%%%%%%%%%%%%%%%%%%%%%%%%%%%%%%%%%%%%%%%%%%%%%%%%%%%%%%%%%%%%%%%%%

It is known that the density perturbation grew much earlier than the time of recombination. If it grew after the recombination time, the density perturbation grown afterward was not enough to make galaxies. For galaxy formation, therefore, DM is needed since proton density perturbation could not grow before the recombination time, but DM could. With DM, the equality point of radiation and matter energy densities can occur much earlier than the recombination time since DM is not prohibited in collapsing by Silk damping \cite{Silkdamp}.
If the WIMP mass and interaction cross section falls in the region allowed by Eq. (\ref{WIMPcond}), it can be CDM. In this talk, we consider CDM as the LSP and in addition axion also. But if the LSP is the only the CDM component, then the LSP mass would give one number for the DM density, which may not be the case. Thus, even if the LSP is contributing to the CDM density, we may need axion to account for the right amount of CDM around us.

\section{Strong CP problem}
Let us start with the discussion on axion's role in the solution of the strong CP problem. Its attractiveness in the strong CP solution is the most attractive feature in advocating axions. The past and future  axion search experiments rely on this theoretical attractiveness.

All the discussion leading to axion started with the discovery of instanton solutions in nonabelian gauge theories \cite{BPRT}, which let to an intrinsic additional parameter $\theta$, the so-called vacuum angle, in nonabelian gauge theories \cite{CDGJR}. In the $\theta$ vacuum, we must consider the P and T (or CP) violating interaction parametrized by $\bar\theta$,
\begin{equation}
{\cal L}= \bar\theta \{F \tilde F\}
\equiv\frac{\bar\theta}{64\pi^2}\epsilon^{\mu\nu\rho\sigma}
F_{\mu\nu}^aF^a_{\rho\sigma}
\end{equation}
where $\bar\theta=\theta_0+\theta_{weak}$ is the final value
including the effects of the electroweak CP violation.  For QCD to become a correct theory, this CP violation by QCD must be
sufficiently suppressed.

It is known that $\bar\theta$ is a physical parameter contributing to the dipole moment of neutron, $d_n$. From the upper bound of $|d_n|<3\times 10^{-26}e$cm \cite{NEDM}, we obtain a bound on $\bar\theta$, $|\bar\theta|<10^{-9}$. The final value of  $|\bar\theta| <10^{-9}$ is perfectly allowed, but the smallness is not explained. The strong CP problem is to understand it more satisfactorily, $\lq\lq$Why is this $\bar\theta$ so small?" So, it is a kind of naturalness problem. There are three explanations for the smallness of $\bar\theta$:
$$
\rm 1.~ Calculable~ \bar\theta, \quad 2.~ Massless~ up~ quark,
\quad 3.~Axion.
$$
We discuss Cases 1 and 2 briefly, and concentrate on Case 3 in the subsequent sections.

\subsubsection{Calculable $\bar\theta$}
 Now, the Nelson-Barr type CP violation is mostly
discussed for the calculable $\bar\theta$ models since it is designed to allow the Kobayashi-Maskawa type weak CP violation at the electroweak scale. It introduces extra fields and corresponding interactions beyond the standard model. For example, vectorlike heavy quarks are introduced at high
energy scales. The scheme is designed such that at low energy the Yukawa couplings are real, which is needed anyway from the beginning to set $\theta_0=0$. This solution is possible with the specific forms for the couplings and their phases and in addition the assumption on VEVs of Higgs doublets \cite{NBarr}.

\subsubsection{Massless up quark}

\begin{figure}[t]
\begin{center}
\begin{picture}(350,210)(0,0)
\Line(50,10)(190,10) \Line(50,10)(50,210)
{\CBox(70,188)(150,70){Yellow}{Yellow}
\CTri(68,188)(84,188)(68,105){White}{White}
\CTri(113,188)(151,188)(151,150){White}{White}
\CTri(99,70)(151,70)(151,135){White}{White}
\CTri(89,70)(68,70)(68,90){White}{White}
 }

 \rText(20,150)[t][l]{$m_d$ [MeV]}
\Line(190,10)(190,210) \Line(50,210)(190,210)
\Line(50,18)(54,18)\Line(50,26)(54,26) \Line(50,34)(54,34)
\Line(50,42)(54,42)\Line(50,50)(57,50) \Text(45,50)[r]{2}
 \Line(50,58)(54,58) \Line(50,66)(54,66) \Line(50,74)(54,74)
 \Line(50,82)(54,82) \Line(50,90)(57,90) \Text(45,90)[r]{4}
 \Line(50,98)(54,98)\Line(50,106)(54,106) \Line(50,114)(54,114)
\Line(50,122)(54,122) \Line(50,130)(57,130) \Text(45,130)[r]{6}
  \Line(50,138)(54,138) \Line(50,146)(54,146)
  \Line(50,154)(54,154) \Line(50,162)(54,162)
  \Line(50,170)(57,170) \Text(45,170)[r]{8} \Line(50,178)(54,178)
\Line(50,186)(54,186)\Line(50,194)(54,194) \Line(50,202)(54,202)
\Text(45,210)[r]{10} \Line(50,18)(54,18)\Line(50,26)(54,26)
\Line(50,34)(54,34) \Line(50,42)(54,42)\Line(50,50)(57,50)
 \Text(45,50)[r]{2}
 \Line(190,18)(186,18)\Line(190,26)(186,26) \Line(190,34)(186,34)
\Line(190,42)(186,42)\Line(190,50)(183,50)
 \Line(190,58)(186,58) \Line(190,66)(186,66) \Line(190,74)(186,74)
 \Line(190,82)(186,82) \Line(190,90)(183,90)
 \Line(190,98)(186,98)\Line(190,106)(186,106) \Line(190,114)(186,114)
\Line(190,122)(186,122) \Line(190,130)(183,130)
  \Line(190,138)(186,138) \Line(190,146)(186,146)
  \Line(190,154)(186,154) \Line(190,162)(186,162)
  \Line(190,170)(183,170)  \Line(190,178)(186,178)
\Line(190,186)(186,186)\Line(190,194)(186,194)
\Line(190,202)(186,202)

\Line(58,10)(58,14)\Line(66,10)(66,14)
\Line(74,10)(74,14)\Line(82,10)(82,14)
 \Line(58,210)(58,206)\Line(66,210)(66,206)
\Line(74,210)(74,206)\Line(82,210)(82,206)
 \Line(90,10)(90,17)\Text(90,0)[c]{ 2}
\Line(98,10)(98,14)\Line(106,10)(106,14)
\Line(114,10)(114,14)\Line(122,10)(122,14)
 \Line(130,10)(130,17)\Text(130,0)[c]{ 4}
 \Line(90,210)(90,203)
\Line(98,210)(98,206)\Line(106,210)(106,206)
\Line(114,210)(114,206)\Line(122,210)(122,206)
 \Line(130,210)(130,203)
\Line(138,10)(138,14)\Line(146,10)(146,14)
\Line(154,10)(154,14)\Line(162,10)(162,14)
 \Line(170,10)(170,17)\Text(170,0)[c]{ 6}
 \Line(130,210)(130,203)
\Line(138,210)(138,206)\Line(146,210)(146,206)
\Line(154,210)(154,206)\Line(162,210)(162,206)
 \Line(170,210)(170,203)
 \Line(178,10)(178,14)\Line(186,10)(186,14)
\Line(178,210)(178,206)\Line(186,210)(186,206)
\Text(130,-15)[c]{$m_u$ [MeV]}

 \Line(90,210)(190,110) \Line(50,10)(190,186)
 \Line(50,110)(150,10) \Line(50,10)(90,210)

\SetWidth{1}
 \Text(96,138)[c]{$\bullet$}\DashLine(96,159)(96,116){3}
 \DashLine(87,138)(105,138){3}

 \Text(110,159)[c]{$\bullet$}\DashLine(110,171)(110,145){3}
 \DashLine(96,159)(124,159){3}

 \Text(129,150)[c]{$\bullet$}\DashLine(129,171)(129,129){3}
 \DashLine(106,150)(154,150){3}

 \Text(127,141)[c]{$\bullet$}\DashLine(127,181)(127,104){3}
 \DashLine(105,141)(147,141){3}

 \SetWidth{1.5}
 \Line(85,188)(112,188)\Line(112,188)(150,150) \Line(150,150)(150,135)
 \Line(150,135)(98,70) \Line(98,70)(90,70)\Line(90,70)(69,90)
 \Line(69,90)(69,105)\Line(69,105)(85,188)

\end{picture}
\end{center}
\caption{The allowed $m_d-m_d$ region bounded by the solid polygon.}
\label{uqmass}
\end{figure}
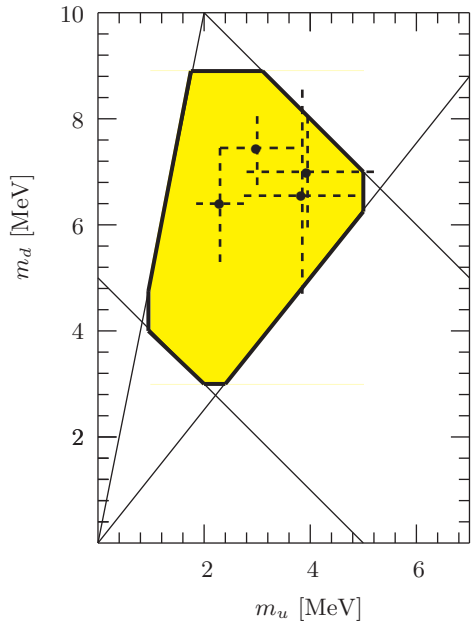

  Suppose that we chiral-transform a quark as $q\to
e^{i\gamma_5\alpha}q$. Then, the QCD Lagrangian changes as
\begin{align}
&\int d^4x [-m_q \bar qq-\bar\theta  \{F \tilde F\} ]\to\nonumber\\
&\quad\int d^4x [-m_q \bar qe^{2i\gamma_5\alpha}q
-(\bar\theta-2\alpha) \{F \tilde F\}]\label{chiraltr}
\end{align}
If $m_q=0$, it is equivalent to changing $\bar\theta\to
\bar\theta-2\alpha$. Thus, there exists a shift symmetry
$\bar\theta\to \bar\theta-2\alpha$. It is known that the tunneling amplitude due to instanton solutions with a zero mass quark vanishes \cite{tHooft}, which implies that the shift symmetry is an exact symmetry. In this case, $\bar\theta$ is not physical, and hence there is no strong CP problem if the lightest quark (i. e. the up quark) is massless. The question for the massless up quark solution is, $\lq\lq$Is the massless up quark phenomenologically viable?" Weinberg's famous up/down quark mass ratio calculation from chiral perturbation theory originally gave 5/9 \cite{Wein79up}, which is very similar to the recent compilation of the light quark masses \cite{Manohar}, $m_u=3\mp 1$ MeV and $m_d=6\pm 1.5$ MeV, which is shown in Fig. \ref{uqmass}. This compilation is convincing enough to rule out the massless up quark possibility
\cite{KapMan}.
%%%%%%%%%%%%%%%%%%%%%%%%%%%%%%%%%%%%%%%%%%%%%%%%%%%%%%%%%%%%%%%%%%%%%%%%%%%%
%%%%%%%%%%%%%%%%%%%%%%%%%%%%%%%%%%%%%%%%%%%%%%%%%%%%%%%%%%%%%%%%%%%%%%%%%%%%
%%%%%%%%%%%%%%%%%%%%%%%%%%%%%%%%%%%%%%%%%%%%%%%%%%%%%%%%%%%%%%%%%%%%%%%%%%%%
For some time the massless up quark possibility was taken seriously
\cite{KapMan}. The reason is that even if the Lagrangian mass for the up quark is zero, the 't Hooft determinental interaction may generate a useful up quark mass for the chiral perturbation. There was a confusion on this issue \cite{Leutwyler,KChoi,lattice}. Now, it is cleared by one of the proponents of the massless up quark that Case 2 is not allowed, without resorting to the lattice result.

\section{Axions}
Peccei and Quinn (PQ) tried to mimic the symmetry  $\bar\theta\to\bar\theta-2\alpha$ of the massless quark case, by considering the full electroweak theory Lagrangian \cite{PQ77}. They found such a symmetry if $H_u$ is coupled only to up-type quarks and $H_d$ couples only to down-type quarks,
$$
{\cal L}=-\bar q_L u_R H_u-\bar q_L d_RH_d-V(H_u,H_d)
+({\rm h.c.})-\bar\theta  \{F \tilde F\}.
$$
Certainly, if we assign the same global charge under the $\gamma_5$ transformation to $H_u$ and $H_d$, $q\to e^{i\gamma_5\alpha}q, H_u\to e^{i\beta}H_u, H_d\to e^{i\beta}H_d$, the flavor independent
part changes to
\begin{align}
{\cal L}\to &-\bar q_Le^{i\gamma_5\alpha} u_R e^{i\beta}H_u
-\bar q_Le^{i\gamma_5\alpha} d_R e^{i\beta}H_d\nonumber\\
&-V(e^{i\beta}H_u,e^{i\beta}H_d)+({\rm h.c.})-(\bar\theta-2\alpha)
\{F \tilde F\}.\label{PQ}
\end{align}
Choosing $\beta=\alpha$ achieves the same kind of only $\bar\theta$ shift as in the massless quark case, which is called the PQ global symmetry U(1)$_{\rm PQ}$. But unlike the massless quark case, here $\bar\theta$ is physical. Even though the coefficient of $\{F\tilde F\}$ changes in the same way in Eqs. (\ref{chiraltr}) and (\ref{PQ}), these two cases differ in that the tunneling amplitude vanishes with a massless quark and does not vanish without a massless quark. Since the tunneling amplitude is not vanishing, physics depends on the value of $\bar\theta$. The $\bar\theta$
dependence of free energy is thus $\sim e^{i\bar\theta}
+e^{-i\bar\theta}\sim \cos{\bar\theta}$. At the classical Lagrangian level, there seems to be no strong CP problem. The phase $\beta$ of the Higgs fields disappears in the classical Lagrangian. But its coupling to $\{F\tilde F\}$ is generated at one loop level, which is the U(1)$_{\rm PQ}$-QCD-QCD anomaly. With this one loop term, the Lagrangian is not invariant under the phase shift symmetry $\beta$. It is explicitly broken and the phase field $\beta$ does not have a flat potential. Weinberg and Wilczek interpreted this phenomenon using the spontaneous symmetry breaking of the global symmetry U(1)$_{\rm PQ}$. The phase $\beta$ turns out to be the Goldstone boson of that spontaneous symmetry breaking of U(1)$_{\rm PQ}$ \cite{WeinWil}. This Goldstone boson is named as {\it axion}. Since $\beta$ (in fact $\alpha$ in Eq. (\ref{PQ})) appears in combination of $\bar\theta-2\beta$, we interpret $\bar\theta$ as the Goldstone field, redefining $\bar\theta-2\beta\to\bar\theta$ and $\frac12 v^2 \partial_\mu\beta\partial^\mu\beta\to \frac12 F_a^2 \partial_\mu\bar\theta\partial^\mu\bar\theta$.
It is said that $\bar\theta$ is made dynamical, but in the theory the component was there from the beginning. The free energy depending on $-\cos\bar\theta$ is the potential for the axion where $a=\bar\theta F_a$. Since it is proportional to $-\cos\bar\theta$, the minimum of the potential is at $\bar\theta=0$ \cite{PQ77,VW},
and the vacuum chooses $\bar\theta=0$. But the weak CP violation shifts $\bar\theta$ a little bit, leading to $\bar\theta\sim O(10^{-17})$ \cite{Randall}. Thus, the axion solution of the strong CP problem is a kind of cosmological solution.

The Peccei-Quinn-Weinberg-Wilczek (PQWW) axion is ruled out soon \cite{PQWWrule}, which was the reason for the popularity of calculable models in 1978 \cite{calcul}. Nowadays, cosmologically considered axions are very light, which arises from the phase of SU(2)$\times$U(1) singlet scalar field $\sigma$. The simplest case is the Kim-Shifman-Vainstein-Zakharov (KSVZ) axion model \cite{KSVZ} which incorporates a heavy quark $Q$ with the following coupling and the resulting chiral symmetry
\begin{align}
{\cal L}=&-\bar Q_LQ_R\sigma+({\rm h.c.})-V(|\sigma|^2)-\bar\theta
\{F \tilde F\},\nonumber\\
{\cal L}\to &-\bar Q_Le^{i\gamma_5\alpha} Q_R e^{i\beta}\sigma
+({\rm h.c.})-V(|\sigma|^2)\nonumber\\
&\quad\quad-(\bar\theta-2\alpha)  \{F \tilde F\}.
\end{align}
Here, Higgs doublets are neutral under U(1)$_{\rm PQ}$. By coupling $\sigma$ to $H_u$ and $H_d$, one can introduce a PQ symmetry also not introducing heavy quarks necessarily, and the resulting axion is called the Dine-Fischler-Srednicki-Zhitnitskii (DFSZ) axion
\cite{DFSZ}. In string models, most probably both heavy quarks and Higgs doublets contribute to the $\sigma$ field couplings.
The VEV of $\sigma$ is much above the electroweak scale
and the axion is a {\it very light axion} \cite{verylight}. The $\sigma$ field may contain very tiny SU(2) doublet components, and hence practically we can consider the axion as the phase of $\sigma$, $\sigma=[(V+\rho)/\sqrt2]e^{ia/F_a} $ with the identification $a\equiv a+2\pi N_{DW}F_a$. Since the domain wall (DW) number appears in the phase of $\sigma$, the magnitude $F_a$ can be somewhat smaller than the singlet VEV $v$,  $F_a=v/N_{DW}$.

\begin{figure}
\begin{center}
\begin{picture}(200,55)(0,0)

\DashLine(105,5)(160,5){3}\LongArrow(160,5)(162,5)
\DashLine(60,15)(60,30){3}\LongArrow(60,30)(60,32)
 \Text(60,42)[c]{$V$}\Text(166,5)[l]{$a$}
\LongArrow(110,-5)(100,-5)\LongArrow(110,-5)(120,-5)
 \Text(110,-13)[c]{$\pi F_a$}
 \Line(37.5,-2.5)(42.5,-2.5)\LongArrow(40,17)(40,11.5)
 \LongArrow(40,-8)(40,-3.5)
 \Text(40,25)[c]{$\Lambda^4_{\rm QCD}$}
 \Text(76.5,10)[c]{$\circ$}

\SetWidth{1} \Photon(10,5)(150,5){-5}{3.5}
\Text(20,1)[c]{\small$\blacktriangledown$}\Text(60,1)[c]{$\bullet$}
 \Text(100,1)[c]{\tiny$\blacksquare$}
\Text(140,1)[c]{\small$\blacktriangledown$}

\end{picture}
\end{center}
\caption{\label{axionexp} The case with $N_{DW}=3$ where three vacua
are distinguished. }\label{Axionpot}
\end{figure}
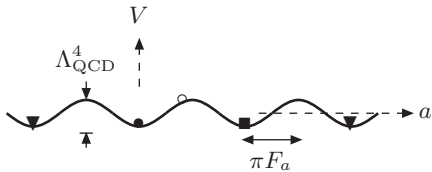

The couplings of $\sigma$ determine the axion interactions. Because the axion $a$ is a Goldstone boson, the axion couplings are of the derivative form except for the anomaly term. In this talk, we will concentrate on this anomaly coupling. Axion is directly related to $\bar\theta$. Its birth was from the PQ symmetry whose spontaneous breaking introduced $a$. Generally, however, we can define $a$ as a
pseudoscalar field without potential terms except the one arising from the gluon anomaly,
\begin{equation}
\frac{a}{F_a}\left\{\frac{g^2}{32\pi^2}F^a_{\mu\nu}
\tilde F^{a\mu\nu} \right\}.\label{gluonanom}
\end{equation}
Then, we note that this kind of nonrenormalizable term can arise in several ways as shown in Table \ref{Models}.
\begin{table}[h]
\begin{center}\begin{tabular}{l|l}
   \hline\\
   [-0.95em]
Axions from &  Order of $F_a$\\[0.2em]
  \hline\\
   [-0.95em]
 String theory& String scale or Planck scale\\[0.3em]
 $M$-theory \cite{ChoiM} & String or the scale of\\
 &\ \  the 11th dimension\\[0.3em]
 Large extra ($n$) dimension& Combination of $M_D$ and $R$\\
 \ \  \cite{extraD}&\ \ [$M_D$ = the fundamental mass,\\
 &\ \
  $R$ = the size of
  extra dimension\\
  &\ \ ({\it cf}. $M_P\simeq M_D(R/M_D)^{n/2}$)]\\[0.3em]
Composite models \cite{composite}& Compositeness scale\\[0.3em]
Renormalizable  theories& The U(1)$_{\rm PQ}$ global symmetry\\
&\ \ breaking scale\\[0.2em]
 \hline
\end{tabular}
\end{center}
\caption{Natural scales of $F_a$.}\label{Models}
\end{table}
For the cases in large extra dimensions, the classification is more complicated since there are many ways to allocate the initial field containing the axion in the bulk and/or branes.
In any case, the essence of the axion solution (wherever it originates in Table \ref{Models}) is that $\langle a\rangle$ seeks $\bar\theta=0$ whatever happened before. In this sense it is a cosmological solution. The potential arising from the anomaly term after integrating out the gluon field is the axion potential. The height of the potential is $O(\Lambda^4_{\rm QCD})$ of the nonabelian gauge interaction, which is shown in Fig. \ref{Axionpot} with the domain wall
number \cite{DWN} $N_{DW}=3$: the bullet, the square and the
triangle denote different vacua.
%%%%%%%%%%%%%%%%%%%%%%%%%%%%%%%%%%%%%%%%%%%%%%%%%%%%%%%%%%%%%%%%%%%%%%%%%%
%%%%%%%%%%%%%%%%%%%%%%%%%%%%%%%%%%%%%%%%%%%%%%%%%%%%%%%%%%%%%%%%%%%%%%%%%
%%%%%%%%%%%%%%%%%%%%%%%%%%%%%%%%%%%%%%%%%%%%%%%%%%%%%%%%%%%%%%%%
Two important properties of axions are: (i) periodic potential with the period $2\pi F_a$, and (ii) the minima are at $a=0, 2\pi F_a, 4\pi F_a, \cdots$. This cosine form of
the potential is determined by considering the axion mixing with pions as
\begin{equation}
V(a)=\frac{Z}{(1+Z)^2}f_\pi^2 m_\pi^2\left[1-\cos\frac{a}{F_a} \right]
\end{equation}
where $Z=m_u/m_d$. Its expansion gives the axion mass
\begin{equation}
m_a\simeq 0.6 \left(\frac{10^7\rm~ GeV}{F_a}\right){\ \rm eV}.
\end{equation}
There can be the axion-photon-photon anomalous coupling of the form $a{\bf E} \cdot{\bf B}$ which can be checked in laboratory, astrophysical and cosmological tests. The old laboratory bound of $F_a>10^4$ GeV has been obtained from meson decays ($J/\Psi\to a\gamma, \Upsilon\to a\gamma, K^+\to a\pi^+$), beam dump experiments ($p(e)N\to aX\to \gamma\gamma X, e^+e^-X$), and nuclear de-excitation ($N^*\to Na\to N\gamma\gamma, Ne^+e^-$)
\cite{axionrev}. The laser induced axion-like particle search has been performed since early 1990s \cite{BFRT,PVLAS}.
%%%%%%%%%%%%%%%%%%%%%%%%%%%%%%%%%%%%%%%%%%%%%%%%%%%%%%%%%%%%%%%%%%%%%%%%%%%
%%%%%%%%%%%%%%%%%%%%%%%%%%%%%%%%%%%%%%%%%%%%%%%%%%%%%%%%%%%%%%%%%%%%%%%%%%%\begin{figure}[b]
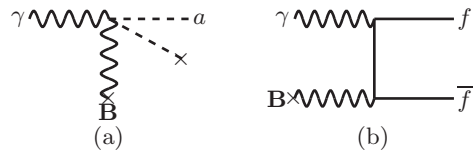
\begin{figure}[h]
\begin{center}
\begin{picture}(400,55)(0,0)
 {\SetWidth{1}
 \Photon(20,45)(50,45){3}{4.5}
 \DashLine(50,45)(80,45){3}\Text(82,45)[l]{$a$}
 \DashLine(50,45)(77.5,30){3}\Text(77.5,30)[c]{$\times$}
 \Photon(50,45)(50,15){3}{4} \Text(50,15)[c]{$\times$}
 \Text(18.5,45)[r]{$\gamma$} \Text(50, 10)[c]{$\bf B$}
 \Text(50,0)[c]{(a)}
 }
 {\SetWidth{1}
 \Photon(120,45)(150,45){3}{5} \Line(150,45)(180,45)
 \Photon(120,15)(150,15){3}{5} \Line(150,45)(150,15)
 \Line(150,15)(180,15)
 \Text(120,15)[c]{$\bf \times$}
 \Text(118.5,45)[r]{$\gamma$} \Text(117.5,15)[r]{$\bf B$}
 \Text(182,45)[l]{$f$}
  \Text(182,15)[l]{$\overline{f}$}
 \Text(150,0)[c]{(b)}
 }
\end{picture}
\end{center}
\caption{Possible processes leading to a vacuum
dichroism.}\label{axionlike}
\end{figure}
%%%%%%%%%%%%%%%%%%%%%%%%%%%%%%%%%%%%%%%%%%%%%%%%%%%%%%%%%%
%%%%%%%%%%%%%%%%%%%%%%%%%%%%%%%%%%%%%%%%%%%%%%%%%%%%%%%%%%
These experiments may find the processes shown in Fig. \ref{axionlike} (axionlike particles in (a) \cite{Moh06} and milli-charged particles in (b) \cite{millicharged}). Here, the polarization of the laser is looked for and actually at present there is no convincing evidence \cite{axlikeexp2} that an anomalous effect was observed, contrary to an earlier confusion \cite{PVLAS}.

\subsection{Axions  from stars}

%%%%%%%%%%%%%%%%%%%%%%%%%%%%%%%%%%%%%%%%%%%%%%%%%%%%%%%%%%%%%%%%%%%%%%%%
%%%%%%%%%%%%%%%%%%%%%%%%%%%%%%%%%%%%%%%%%%%%%%%%%%%%%%%%%%%%%%%%%%%%%%%%
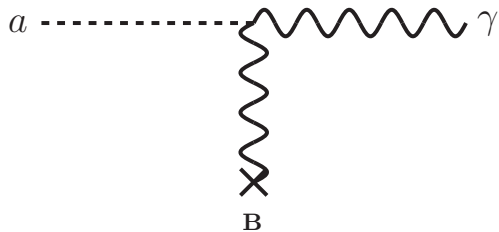
\begin{figure}
\begin{center}
\begin{picture}(400,100)(0,20)
\SetWidth{1.5} \Text(35,90)[r]{\Large $a$} \Text(205,90)[l]{\Large
$\gamma$}
 \DashLine(40,90)(120,90){3}
\Photon(120,90)(200,90){5}{5}
 \Photon(120,30)(120,90){-5}{4}
 \Line(125,35)(115,25)\Line(125,25)(115,35)
 \Text(120,15)[c]{$\bf B$}

\end{picture}
\end{center}
\caption{The axion$\leftrightarrow$photon conversion in a magnetic field. }
\end{figure}
%%%%%%%%%%%%%%%%%%%%%%%%%%%%%%%%%%%%%%%%%%%%%%%%%%%%%%%%%%%%%%%%%%%%%%%%%
We use the axion couplings to $e, p, n,$ and photon to study the core evolution of a star.  The important process is the Primakoff process for which the following coupling $c_{a\gamma\gamma}$ is assumed,
\begin{align}
{\cal L}&=-c_{a\gamma\gamma}\frac{a}{F_a}\{F_{\rm em}\tilde F_{\rm em}\},\ \
c_{a\gamma\gamma} =\bar c_{a\gamma\gamma}-1.93\nonumber\\
& \bar c_{a\gamma\gamma}={\rm Tr} Q^2_{\rm em}|_{E\gg M_Z}.
\end{align}
Axion helioscopes of Tokyo \cite{Tokyo} and CAST (CERN axion solar telescope) \cite{CAST} and also the geomagnetic conversion use this photon coupling.

\begin{figure}[t]
\centering \epsfig{figure=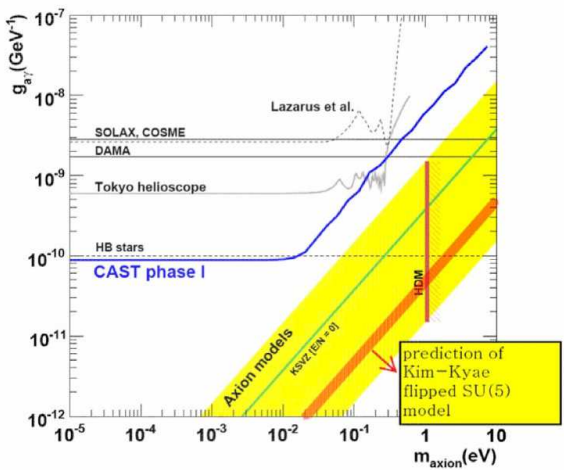, width=9.2cm}
\caption{The CAST bound with other constraints. Here, we show some field theoretic models and the prediction from the string $Z_{12-I}$ model [81]. }\label{CASTexp}
\end{figure}

In the hot plasma in stars, axions once produced most probably
escape the core of the star and take out energy. This contributes to the energy loss mechanism of star and should not dominate the luminocity. From the Primakoff process which is present in any model, we obtain $F_a>10^7$ GeV except the $m_a>200$ keV region for which $a$ is too heavy to be produced in the core of a star. From the nucleon-nucleon-$a$ coupling, SN1987A gave a strong bound $F_a>0.6\times 10^9$ GeV \cite{RafTur} with the correct nucleon state taken into account in \cite{ChoiKang}. The recent refined analysis gives $F_a>10^{10}$ GeV \cite{Raffelt06}. For the DFSZ axion, the $aee$ coupling is present and the Compton-like scattering
($\gamma e\to ae$) in the core gave the bound $g_{aee}<1.3\times 10^{-13}$ \cite{Raffelt06}.

Laboratory experiments can perform more than just the energy loss mechanism in the core of a star. The early Tokyo experiment \cite{Tokyo} could not give a more stringent bound than the supernova limit, but the CAST could compete with the supernova bound. The CAST result is shown in Fig. \ref{CASTexp} together with other bounds \cite{CAST}. In this figure, hypothetical field theoretic models are schematically shown also.

\subsection{Axions in the universe}

\begin{figure}[h]
\begin{center}
\begin{picture}(400,38)(0,25)
{\SetWidth{1.5}
\DashLine(20,20)(220,20){3}
\Photon(20,30.5)(220,30.5){-10}{0.5}
{\CCirc(210,31.8){1.5}{Red}{Red}}
%\LongArrow(120,25)(135,25)
 }
 \CBox(120,24)(130,26){Blue}{Blue}
 \CTri(129,22)(129,28)(134,25){Blue}{Blue}
  \LongArrow(120,35)(208,35)
  \Text(160,40)[c]{$O(F_a)$}
\end{picture}
\end{center}
\caption{The almost flat axion potential. The misalignment angle is expected to be of order 1 but can also be very small as shown by the thick blue arrow.}\label{flataxionp}
\end{figure}
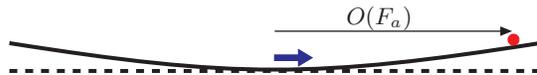

 The potential of the very light axion is of almost
flat. Therefore, a chosen vacuum point (the red bullet in Fig.
\ref{flataxionp}) stays there for a long time, and starts to oscillate when the Hubble time $H^{-1}$ is comparable to the oscillation period (the inverse axion mass), $H<m_a$.  This occurs when the temperature of the universe is about 1 GeV \cite{PWW}. The DW problem in the standard Big Bang cosmology \cite{SikDW} disappears when the reheating temperature after inflation is below the PQ symmetry breaking scale. Since the reheating temperature is $T_{\rm RH}<10^9$ GeV or $10^7$ GeV in some models \cite{EKN}, here we will not worry about the DW problem any more.
%%%%%%%%%%%%%%%%%%%%%%%%%%%%%%%%%%%%%%%%%%%%%%%%%%%%%%%%%%%%%%%%%%%%%%%%
%%%%%%%%%%%%%%%%%%%%%%%%%%%%%%%%%%%%%%%%%%%%%%%%%%%%%%%%%%%%%%%%%%%%%%%%
%%%%%%%%%%%%%%%%%%%%%%%%%%%%%%%%%%%%%%%%%%%%%%%%%%%%%%%%%%%%%%%%%%%%%%%%%

The axion is created at $T=F_a$, but the universe $\langle a\rangle$ does not begin to roll until $H=m_a$, i.e. at $T=1$ GeV. From then, the classical field  $\langle a\rangle$  starts to oscillate. The energy density behaves like that in the harmonic oscillator $m_a^2F_a^2$ which is proportional to the axion mass times the number density. Thus, its behavior is like that of CDM, which is the reason that the axion DM is CDM even though its mass is very small and its interaction strength is much weaker than $\lq\lq$weak". For a recent review, see \cite{Asztalos}. The axion CDM energy fraction is
\begin{equation}
\Omega_a=\frac12 \left(\frac{0.6\times 10^{-5}{\rm eV}}{m_a}
\right)^{7/6}
\end{equation}
If $F_a$ is large($> 10^{12}$ GeV), then the axion energy density dominates the universe. Summarizing the astro and cosmological constraints, we take the axion $F_a$ window as,
\begin{equation}
10^{10}\ {\rm GeV} \le
F_a\le 10^{12}\ \rm GeV.
\end{equation}
But there is an anthropic argument even beyond
$F_a> 10^{12}$ GeV.

\begin{figure}[t]
\begin{center}
\begin{picture}(350,230)(0,-10)
\SetScale{0.8}
\Line(50,10)(300,10) \Line(50,10)(50,260)
 \rText(10,180)[t][l]{$Q$, Scalar fluctuation amplitude}
 \Text(140,-12)[c]{$\xi$, Matter density per CMB photon [eV]}
\Line(300,10)(300,260)\Line(50,260)(300,260)

{\CBox(51,175)(145,89){Yellow}{Yellow}
\CTri(51,175)(145,175)(145,127){White}{White}
\CTri(145,89)(51,89)(51,122){White}{White}
}

\Line(50,27.5)(55,27.5)\Line(295,27.5)(300,27.5)
\Line(50,35)(57,35)\Line(50,52.5)(55,52.5)\Line(295,52.5)(300,52.5)
\Text(38,30)[r]{\small $10^{-9}$}\Line(293,35)(300,35)
\Line(50,60)(57,60)\Line(50,77.5)(55,77.5)\Line(295,77.5)(300,77.5)
\Text(38,50)[r]{\small $10^{-8}$}\Line(293,60)(300,60) \Line(50,85)(57,85)
\Text(38,70)[r]{\small $10^{-7}$}\Line(293,85)(300,85)
\Line(50,102.5)(55,102.5)\Line(295,102.5)(300,102.5)
\Line(50,110)(57,110)\Line(50,127.5)(55,127.5)
\Line(295,127.5)(300,127.5)
\Text(38,90)[r]{\small $10^{-6}$}\Line(293,110)(300,110)
\Line(50,135)(57,135)\Line(50,152.5)(55,152.5)
\Line(295,152.5)(300,152.5)
\Text(38,110)[r]{\small $10^{-5}$}\Line(293,135)(300,135)
\Line(50,160)(57,160)\Line(50,177.5)(55,177.5)
\Line(295,177.5)(300,177.5)
\Text(38,130)[r]{\small $10^{-4}$}\Line(293,160)(300,160)
\Line(50,185)(57,185)\Line(50,202.5)(55,202.5)
\Line(295,202.5)(300,202.5)
\Text(38,150)[r]{\small $10^{-3}$}\Line(293,185)(300,185)
\Line(50,210)(57,210)\Line(50,227.5)(55,227.5)
\Line(295,227.5)(300,227.5)
\Text(38,170)[r]{\small $10^{-2}$}\Line(293,210)(300,210)
\Line(50,235)(57,235)\Line(50,252.5)(55,252.5)
\Line(295,252.5)(300,252.5)
\Text(38,190)[r]{\small $0.1$}\Line(293,235)(300,235)

\Text(40,0)[c]{\small 1} \Line(65,10)(65,13)\Line(65,260)(65,257)
\Line(74,10)(74,13)\Line(74,260)(74,257)\Line(80,10)(80,13)
\Line(89,10)(89,13)\Line(89,260)(89,257)
\Line(92.3,10)(92.3,13)\Line(92.3,260)(92.3,257)
\Line(95.2,10)(95.2,13)\Line(95.2,260)(95.2,257)
\Line(97.7,10)(97.7,13)\Line(97.7,260)(97.7,257)
\Line(80,260)(80,257)\Line(85,10)(85,15)\Line(85,260)(85,255)

\Line(100,10)(100,17) \Line(115,10)(115,13)
\Line(115,260)(115,257)\Line(124,10)(124,13)\Line(124,260)(124,257)
\Line(130,10)(130,13)\Line(130,260)(130,257)
\Line(135,10)(135,15)\Text(80,0)[c]{\small $10^1$}
\Line(139,10)(139,13)\Line(139,260)(139,257)
\Line(142.3,10)(142.3,13)\Line(142.3,260)(142.3,257)
\Line(145.2,10)(145.2,13)\Line(145.2,260)(145.2,257)
\Line(147.7,10)(147.7,13)\Line(147.7,260)(147.7,257)
\Line(100,260)(100,253)\Line(135,260)(135,255)

\Line(150,10)(150,17)\Line(165,10)(165,13)\Line(165,260)(165,257)
\Line(174,10)(174,13)\Line(174,260)(174,257)
\Line(180,10)(180,13)\Line(180,260)(180,257)
\Line(189,10)(189,13)\Line(189,260)(189,257)
\Line(192.3,10)(192.3,13)\Line(192.3,260)(192.3,257)
\Line(195.2,10)(195.2,13)\Line(195.2,260)(195.2,257)
\Line(197.7,10)(197.7,13)\Line(197.7,260)(197.7,257)
\Line(185,10)(185,15)\Text(120,0)[c]{\small $10^2$}\Line(150,260)(150,253)
\Line(185,260)(185,255)

 \Line(200,10)(200,17) \Line(215,10)(215,13)\Line(215,260)(215,257)
 \Line(224,10)(224,13)\Line(224,260)(224,257)
 \Line(230,10)(230,13)\Line(230,260)(230,257)
 \Line(235,10)(235,15)\Line(239,10)(239,13)\Line(239,260)(239,257)
 \Line(242.3,10)(242.3,13)\Line(242.3,260)(242.3,257)
\Line(245.2,10)(245.2,13)\Line(245.2,260)(245.2,257)
\Line(247.7,10)(247.7,13)\Line(247.7,260)(247.7,257)
\Text(160,0)[c]{\small $10^3$}\Line(200,260)(200,253)
\Line(235,260)(235,255)

 \Line(250,10)(250,17) \Line(265,10)(265,13)\Line(265,260)(265,257)
 \Line(274,10)(274,13)\Line(274,260)(274,257)
 \Line(280,10)(280,13)\Line(280,260)(280,257)
 \Line(289,10)(289,13)\Line(289,260)(289,257)
 \Line(292.3,10)(292.3,13)\Line(292.3,260)(292.3,257)
\Line(295.2,10)(295.2,13)\Line(295.2,260)(295.2,257)
\Line(297.7,10)(297.7,13)\Line(297.7,260)(297.7,257)
\Text(200,0)[c]{\small $10^4$}\Line(250,260)(250,253)

\Line(285,10)(285,15)\Line(285,260)(285,255)
\Text(240,0)[c]{\small $10^5$}

%%%%%%%%Black hole trouble
\DashLine(50,235)(300,235){3}\Text(80,196)[l]{Black hole trouble}
\DashLine(50,225)(300,100){3}\Text(100,175)[l]{Black hole trouble?}
 \Text(100,165)[l]{(Nonlinear before decoupling)}
\DashLine(50,60)(300,60){3}\Text(90,30)[l]{Line cooling freezeout}
\DashLine(50,122)(228,60){3}\Text(55,60)[l]{Cooling trouble}
\DashLine(50,175)(280,60){3}\Text(80,135)[l]{Too close}
                            \Text(90,125)[l]{encounters?}
\DashLine(145,127)(145,89){3}
 \PText(160,105)(-25)[l]{No fragmentation}
 \Text(64,116)[c]{\Large$\star$}
%\Text(80,142.5)[c]{\Large$\star$}

 \SetWidth{1.5} \Line(50,175)(145,127)\Line(145,127)(145,89)
 \Line(145,89)(50,122)

 \end{picture}
\end{center}
\caption{$Q$ versus $\xi$. The anthropically allowed region is
 shaded yellow, and the $\star$ is at $(4\ {\rm eV}, 2\times
10^{-5})$.}\label{Anthropic}
\end{figure}
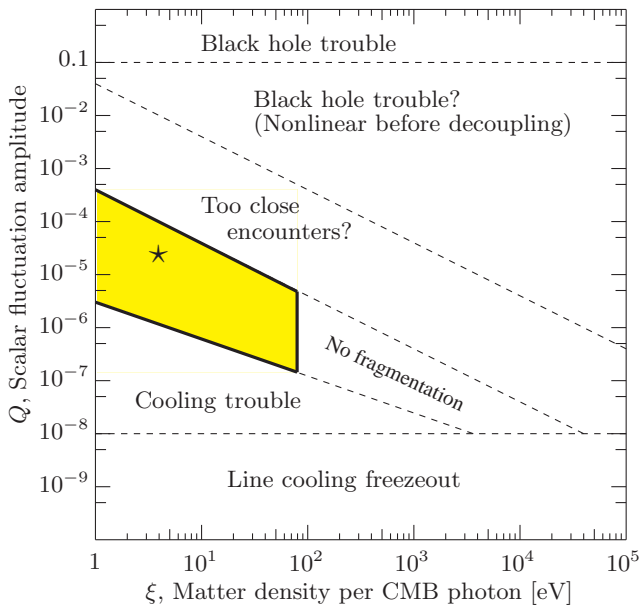

The first anthropic argument on axion was given by Pi in the new inflationary model \cite{Pi} and the recent more refined version is given by Tegmark et. al. \cite{Tegmark}. The homogeneous axion field value (with $a\to -a$ symmetry) right after inflation can take any value between 0 and $\pi F_a$ or $\bar\theta_{\rm mis}=[0,\pi]$. So $\Omega_a$ may be at the required value by an appropriate initial misalignment angle for  $F_a> 10^{12}$ GeV  by taking the needed $\bar\theta_{\rm mis}$ in the new, chaotic or hybrid inflationary scenarios. Tegmark et. al. studied the landscape scenario for 31 dimensionless parameters and some dimensionful parameters with which habitable planets are considered for the assumed nuclear physics parameters. They argue that the prior probability function is calculable for axion models, which is rather obvious. For the axion
with the almost flat potential, it is almost equally probable for $\langle a\rangle$ to sit anywhere in the region $\langle a\rangle=[0,F_a]$ since its location does not affect the other anthropic arguments conspicuously. For axion, one relevant figure for our purpose is scalar fluctuation $Q\simeq\delta\rho/\rho$ vs $\xi$ (matter density per CMB photon), which is shown in Fig.
\ref{Anthropic}.
%%%%%%%%%%%%%%%%%%%%%%%%%%%%%%%%%%%%%%%%%%%%%%%%%%%%%%%%%%%%%%%%%%%%%%%%%%%%
%%%%%%%%%%%%%%%%%%%%%%%%%%%%%%%%%%%%%%%%%%%%%%%%%%%%%%%%%%%%%%%%%%%%%%%%%%%%
%%%%%%%%%%%%%%%%%%%%%%%%%%%%%%%%%%%%%%%%%%%%%%%%%%%%%%%%%%%%%%%%%%%%%%%%%%%%

If a WIMP is the sole candidate for CDM, one obtains just one number for $\delta\rho/\rho$. This may not fit to the observed point ($\star$) in Fig. \ref{Anthropic}. So, we may need the CDM favored WIMP and in addition the axion with $F_a>10^{12}$ GeV. This fits very well to the anthropic principle since any extra $\delta\rho/\rho$ can be provided by axion with an appropriate $\bar\theta_{\rm mis}$ with $F_a>10^{12}$ GeV. Namely, WIMPs may be dominantly the CDM, and the rest amount of DM is provided by axions using some of the anthropic arguments. The Compton wave length of micro eV axion is roughly 20 cm. So, the optimal size for detecting a micro eV axion is of dimension (20 cm)$^3$. The optimal dimension of the detector depends on the mass of axion, which is the key difficulty in the cosmic axion search.

\begin{figure}[t]
\begin{center}
\begin{picture}(400,250)(-10,-55)
\SetScale{0.65}
\Line(30,-50)(350,-50) \Line(30,270)(350,270) \Line(30,-50)(30,270)
\Line(350,-50)(350,270) \Text(170,-57)[r]{%\Large
$m_{\rm axion}$[eV]}
 \rText(-5,160)[t][l]{%\Large
 $|f_{a\gamma\gamma}|=|c_{a\gamma\gamma}|\cdot
 1.93\cdot 10^{-10}m_{a,\rm eV}$ [GeV$^{-1}$]}
\Text(18,-40)[c]{$10^{-6}$} \Line(110,270)(110,263)
\Line(110,-50)(110,-43)\Text(72,-40)[c]{$10^{-5}$}
  \Line(190,-50)(190,-43) \Line(190,270)(190,263)
  \Text(125,-40)[c]{$10^{-4}$}
   \Line(270,-50)(270,-43) \Line(270,270)(270,263)
    \Text(177,-40)[c]{$10^{-3}$}
  \Text(225,-40)[c]{$10^{-2}$}

  %%RBF
 \CBox(81.5,270)(85.9,57.9){Blue}{Blue}
 \CBox(85.9,270)(89.2,74.6){Blue}{Blue}
 \CBox(100,270)(103.5,108.2){Blue}{Blue}
 \CBox(103.5,270)(107.9,114){Blue}{Blue}
 \CBox(107.9,270)(112,91.7){Blue}{Blue}
 \CBox(114,270)(126.5,150.4){Blue}{Blue}
 \Text(52,182)[l]{\small RBF: blue}

%%U of Florida
 \CBox(97.5,270)(100,92){Blue}{Blue}
 \CBox(99.5,270)(100,69.3){Red}{Red}
 %\CBox(97.5,270)(100,92){Red}{Red}
 \CBox(99.5,270)(100,100){Red}{Red}
   \CBox(89.2,270)(97.1,67){Red}{Red}
 \CBox(89.2,44.1)(92.3,67){Red}{Red}
 \Text(52,190)[l]{\small Florida: red}

%%ADMX
 \CBox(53.6,270)(56.9,-74){Brown}{Brown}
 \rText(47,100)[r][l]{ADMX exp. (HR)}

%%Future ADMX
 {\SetColor{Brown}\SetWidth{1.5}
 \DashLine(190,270)(190,30){3}\DashLine(190,30)(90,-70){3}
 \DashArrowLine(90,-70)(85,-75){3}
 \rText(132,130)[r][l]{Future ADMX}
 }
%%%%%%%%%%%%%%%%%%%%%%%%%%%%%%%%%%%%%%%%%%%%%%%%%%%%%%%%%

 \Line(30,30)(37,30)\Line(350,30)(343,30)
 \rText(10,21)[c][l]{\small $10^{-14}$}
\Line(30,110)(37,110)\Line(350,110)(343,110)
 \rText(10,73)[c][l]{\small $10^{-13}$}
 \Line(30,190)(37,190)\Line(350,190)(343,190)
 \rText(10,125)[c][l]{\small $10^{-12}$}
 \Line(30,270)(37,270)\Line(350,270)(343,270)
 \rText(10,175)[c][l]{\small $10^{-11}$}%%%%%%%%%%%%%%%%%%%%%%%%%%%%%%%%%%%%%%%

 \Line(30,-25.9)(34,-25.9) \Line(350,-25.9)(346,-25.9)
 \Line(30,-11.8)(34,-11.8) \Line(350,-11.8)(346,-11.8)
 \Line(30,-1.8)(34,-1.8) \Line(350,-1.8)(346,-1.8)
 \Line(30,5.9)(36,5.9) \Line(350,5.9)(344,5.9)
   \Line(30,12.3)(34,12.3) \Line(350,12.3)(346,12.3)
  \Line(30,17.6)(34,17.6) \Line(350,17.6)(346,17.6)
 \Line(30,22.2)(34,22.2) \Line(350,22.2)(346,22.2)
 \Line(30,26.3)(34,26.3) \Line(350,26.3)(346,26.3)

 \Line(30,54.1)(34,54.1) \Line(350,54.1)(346,54.1)
 \Line(30,68.2)(34,68.2) \Line(350,68.2)(346,68.2)
 \Line(30,78.2)(34,78.2) \Line(350,78.2)(346,78.2)
 \Line(30,85.9)(36,85.9) \Line(350,85.9)(344,85.9)
  \Line(30,92.3)(34,92.3) \Line(350,92.3)(346,92.3)
  \Line(30,97.6)(34,97.6) \Line(350,97.6)(346,97.6)
 \Line(30,102.2)(34,102.2) \Line(350,102.2)(346,102.2)
 \Line(30,106.3)(34,106.3) \Line(350,106.3)(346,106.3)

 \Line(30,134.1)(34,134.1) \Line(350,134.1)(346,134.1)
 \Line(30,148.2)(34,148.2) \Line(350,148.2)(346,148.2)
 \Line(30,158.2)(34,158.2) \Line(350,158.2)(346,158.2)
 \Line(30,165.9)(36,165.9) \Line(350,165.9)(344,165.9)
  \Line(30,172.3)(34,172.3) \Line(350,172.3)(346,172.3)
  \Line(30,177.6)(34,177.6) \Line(350,177.6)(346,177.6)
  \Line(30,182.2)(34,182.2) \Line(350,182.2)(346,182.2)
 \Line(30,186.3)(34,186.3) \Line(350,186.3)(346,186.3)

 \Line(30,214.1)(34,214.1) \Line(350,214.1)(346,214.1)
 \Line(30,228.2)(34,228.2) \Line(350,228.2)(346,228.2)
 \Line(30,238.2)(34,238.2) \Line(350,238.2)(346,238.2)
 \Line(30,245.9)(36,245.9) \Line(350,245.9)(344,245.9)
  \Line(30,252.3)(34,252.3) \Line(350,252.3)(346,252.3)
  \Line(30,257.6)(34,257.6) \Line(350,257.6)(346,257.6)
  \Line(30,262.2)(34,262.2) \Line(350,262.2)(346,262.2)
 \Line(30,266.3)(34,266.3) \Line(350,266.3)(346,266.3)

 \Line(85.9,-50)(85.9,-45)\Line(85.9,270)(85.9,265)
 \Line(165.9,-50)(165.9,-45)\Line(165.9,270)(165.9,265)
 \Line(245.9,-50)(245.9,-45)\Line(245.9,270)(245.9,265)
 \Line(325.9,-50)(325.9,-45)\Line(325.9,270)(325.9,265)

 {\SetWidth{1}\Line(87,-50)(350,213)
 \Line(121,-50)(350,179)
 \PText(220,100)(45)[l]{KSVZ: e_Q=0}
 \PText(230,75)(45)[l]{DFSZ: d^c unification}
  \Line(350,187.5)(355,187.5)\Text(234,122.5)[l]{$1$}
\Line(350,219)(355,219)\Text(233,143)[l]{$\frac{-1}3$}
 \Text(228,187)[c]{$e_Q=$}
 }

 {\SetColor{OliveGreen}\SetWidth{1}
 \DashLine(350,143.2)(156.8,-50){3}}
 \PText(230,37)(45)[l]{A flipped-SU(5) model}

  \end{picture}
\end{center}
 \caption{ The
bounds on cosmic axion searches with some theoretical expectations.
}\label{CosmicAxion}
\end{figure}
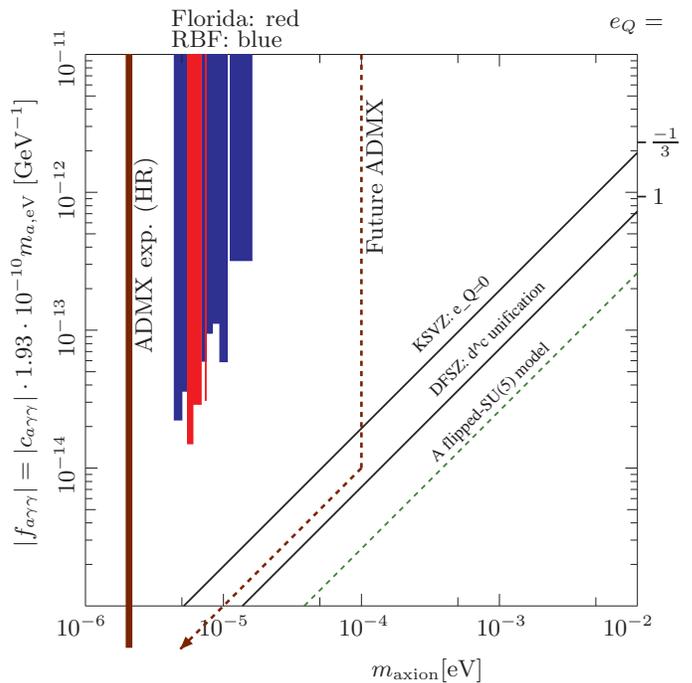

If axion is the CDM component of the universe, then they can be detected even though it may be very difficult. The feeble axion coupling can be compensated by a huge number of axions, since the number density is $\sim F_a^2$ and the cross section is $\sim 1/F_a^2$. So, there is a hope to detect cosmic axions, which has been realized by Sikivie's cavity detector \cite{SikDet}. To detect axion masses in the region $10^{-6}$ eV, one needs low temperature cavity with dimension $O(>10^4~\rm cm^3)$ and the magnetic field strength of $O(10~\rm Tesla)$. The current status of cosmic axion search is shown in Fig. \ref{CosmicAxion}.

%%%%%%%%%%%%%%%%%%%%%%%%%%%%%%%%%%%%%%%%%%%%%%%%%%%%%%%%%%%%%%%%%%%%%%
%%%%%%%%%%%%%%%%%%%%%%%%%%%%%%%%%%%%%%%%%%%%%%%%%%%%%%%%%%%%%%%%%%%%%%
%%%%%%%%%%%%%%%%%%%%%%%%%%%%%%%%%%%%%%%%%%%%%%%%%%%%%%%%%%%%%%%%%%%%%%%%
%%%%%%%%%%%%%%%%%%%%%%%%%%%%%%%%%%%%%%%%%%%%%%%%%%%%%%%%%%%%%%%%%%%%%%%%

\section{SUSY extension and axino}
Not to have too many  thermally produced gravitinos after inflation, the reheating temperature must be bounded \cite{EKN}: $T_{\rm RH}<10^9$ GeV (old bound) or $T_{\rm RH}<10^7$ GeV (new bound if $M_{\rm gluino}<m_{3/2}$). Thus, in SUSY theories  we must consider a relatively small reheating temperature. The axion solution of the strong CP problem with SUSY implies the superpartner of axion, {\it axino}, with a low reheating temperature. The neutralino LSP seems the most attractive candidate for DM simply because the TeV order SUSY breaking scale introduces the LSP as a WIMP. This scenario needs an exact or an effective R-parity for proton to be sufficiently long lived.

Axion's scalar partner, saxion, can also affect the cosmic evolution, but its effect is not so dramatic as the effect of axino \cite{saxion}.

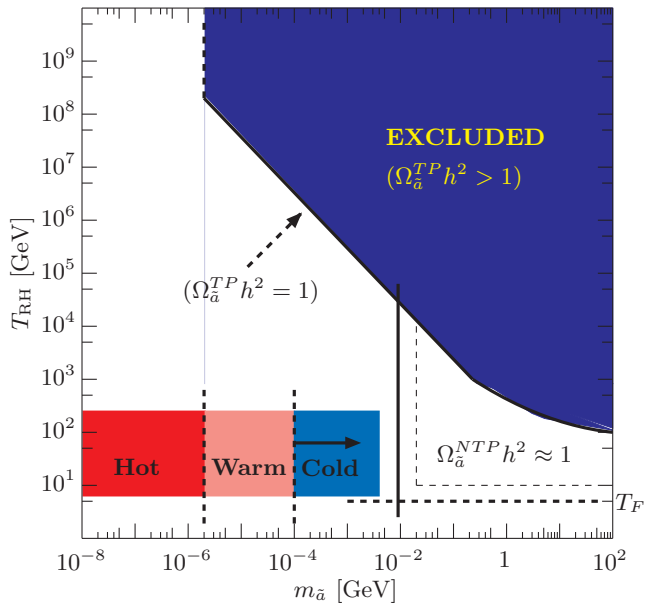
\begin{figure}[t]
\begin{center}
\begin{picture}(350,215)(10,0)
\SetScale{0.8}
   \CBox(50.5,70)(108,30){Red}{Red}
  \CBox(108,30)(150,70){Salmon}{Salmon}
  \CBox(150,30)(190,70){NavyBlue}{NavyBlue}
  \CBox(108,260)(300,83){Blue}{Blue}
  \CTri(108,218)(108,82)(236.5,82){White}{White}
  \CTri(237,83)(300,83)(300,62){Blue}{Blue}
  \CTri(237,85)(265,67)(300,61){Blue}{Blue}
  \CTri(240,68)(260,66)(280,63){White}{White}
  \CTri(255,70)(263,68)(275,63){White}{White}

\Line(50,10)(300,10) \Line(50,10)(50,260)
 \rText(16,130)[t][l]{$T_{\rm RH}$ [GeV]}
 \Text(140,-12)[c]{$m_{\tilde a}$ [GeV]}
\Line(300,10)(300,260)\Line(50,260)(300,260)

\Line(50,27.5)(55,27.5)\Line(295,27.5)(300,27.5)
\Line(50,35)(57,35)\Line(50,52.5)(55,52.5)\Line(295,52.5)(300,52.5)
\Text(38,30)[r]{$10^{1}$}\Line(293,35)(300,35)
\Line(50,60)(57,60)\Line(50,77.5)(55,77.5)\Line(295,77.5)(300,77.5)
\Text(38,50)[r]{$10^{2}$}\Line(293,60)(300,60) \Line(50,85)(57,85)
\Text(38,70)[r]{$10^{3}$}\Line(293,85)(300,85)
\Line(50,102.5)(55,102.5)\Line(295,102.5)(300,102.5)
\Line(50,110)(57,110)\Line(50,127.5)(55,127.5)
\Line(295,127.5)(300,127.5)
\Text(38,90)[r]{$10^{4}$}\Line(293,110)(300,110)
\Line(50,135)(57,135)\Line(50,152.5)(55,152.5)
\Line(295,152.5)(300,152.5)
\Text(38,110)[r]{$10^{5}$}\Line(293,135)(300,135)
\Line(50,160)(57,160)\Line(50,177.5)(55,177.5)
\Line(295,177.5)(300,177.5)
\Text(38,130)[r]{$10^{6}$}\Line(293,160)(300,160)
\Line(50,185)(57,185)\Line(50,202.5)(55,202.5)
\Line(295,202.5)(300,202.5)
\Text(38,150)[r]{$10^{7}$}\Line(293,185)(300,185)
\Line(50,210)(57,210)\Line(50,227.5)(55,227.5)
\Line(295,227.5)(300,227.5)
\Text(38,170)[r]{$10^{8}$}\Line(293,210)(300,210)
\Line(50,235)(57,235)\Line(50,252.5)(55,252.5)
\Line(295,252.5)(300,252.5)
\Text(38,190)[r]{$10^9$}\Line(293,235)(300,235)

 \Line(57.5,10)(57.5,13)\Line(57.5,260)(57.5,257)
 \Line(62,10)(62,13)\Line(62,260)(62,257)
\Line(65,10)(65,13)\Line(65,260)(65,257)

\Text(40,0)[c]{$10^{-8}$}
\Line(67.5,10)(67.5,15)\Line(67.5,260)(67.5,255)
\Line(75,10)(75,17)\Line(75,260)(75,253)
\Line(90,10)(90,13)\Line(90,260)(90,257)
\Line(82.5,10)(82.5,13)\Line(82.5,260)(82.5,257)
\Line(87,10)(87,13)\Line(87,260)(87,257)
\Line(92.5,10)(92.5,15)\Line(92.5,260)(92.5,255)
\Line(100,10)(100,17)\Line(100,260)(100,253)
\Text(80,0)[c]{$10^{-6}$}
\Line(107.5,10)(107.5,13)\Line(107.5,260)(107.5,257)
\Line(112,10)(112,13)\Line(112,260)(112,257)
\Line(115,10)(115,13)\Line(115,260)(115,257)
\Line(117.5,10)(117.5,15)\Line(117.5,260)(117.5,255)
\Line(125,10)(125,17)\Line(125,260)(125,253)
\Line(132.5,10)(132.5,13)\Line(132.5,260)(132.5,257)
\Line(137,10)(137,13)\Line(137,260)(137,257)
\Line(140,10)(140,13)\Line(140,260)(140,257)
\Line(142.5,10)(142.5,15)\Line(142.5,260)(142.5,255)
\Line(150,10)(150,17)\Line(150,260)(150,253)
\Text(120,0)[c]{$10^{-4}$}
\Line(157.5,10)(157.5,13)\Line(157.5,260)(157.5,257)
\Line(162,10)(162,13)\Line(162,260)(162,257)
\Line(165,10)(165,13)\Line(165,260)(165,257)
\Line(167.5,10)(167.5,15)\Line(167.5,260)(167.5,255)
\Line(175,10)(175,17)\Line(175,260)(175,253)
\Line(182.5,10)(182.5,13)\Line(182.5,260)(182.5,257)
\Line(187,10)(187,13)\Line(187,260)(187,257)
\Line(190,10)(190,13)\Line(190,260)(190,257)
\Line(192.5,10)(192.5,15)\Line(192.5,260)(192.5,255)
\Line(200,10)(200,17)\Line(200,260)(200,253)
\Text(160,0)[c]{$10^{-2}$}
\Line(207.5,10)(207.5,13)\Line(207.5,260)(207.5,257)
\Line(212,10)(212,13)\Line(212,260)(212,257)
\Line(215,10)(215,13)\Line(215,260)(215,257)
\Line(217.5,10)(217.5,15)\Line(217.5,260)(217.5,255)
\Line(225,10)(225,17)\Line(225,260)(225,253)
\Line(232.5,10)(232.5,13)\Line(232.5,260)(232.5,257)
\Line(237,10)(237,13)\Line(237,260)(237,257)
\Line(240,10)(240,13)\Line(240,260)(240,257)
\Line(242.5,10)(242.5,15)\Line(242.5,260)(242.5,255)
\Line(250,10)(250,17)\Line(250,260)(250,253)\Text(200,0)[c]{$1$}
\Line(257.5,10)(257.5,13)\Line(257.5,260)(257.5,257)
 \Line(262,10)(262,13)\Line(262,260)(262,257)
\Line(265,10)(265,13)\Line(265,260)(265,257)
\Line(267.5,10)(267.5,15)\Line(267.5,260)(267.5,255)
\Line(269.5,10)(269.5,13)\Line(269.5,260)(269.5,257)
\Line(271.1,10)(271.1,13)\Line(271.1,260)(271.1,257)
\Line(272.6,10)(272.6,13)\Line(272.6,260)(272.6,257)
\Line(273.9,10)(273.9,13)\Line(273.9,260)(273.9,257)
\Line(275,10)(275,17)\Line(275,260)(275,253) \Text(240,0)[c]{$10^2$}
\Line(282.5,10)(282.5,13)\Line(282.5,260)(282.5,257)
\Line(287,10)(287,13)\Line(287,260)(287,257)
\Line(290,10)(290,13)\Line(290,260)(290,257)
\Line(292.5,10)(292.5,15)\Line(292.5,260)(292.5,255)
\Line(294.5,10)(294.5,13)\Line(294.5,260)(294.5,257)
\Line(296.1,10)(296.1,13)\Line(296.1,260)(296.1,257)
\Line(297.6,10)(297.6,13)\Line(297.6,260)(297.6,257)
\Line(298.9,10)(298.9,13)\Line(298.9,260)(298.9,257)

%%%%%%%%Hot
{\SetWidth{1.5}
\DashLine(107.5,17)(107.5,80){3}
\DashLine(107.5,253)(107.5,217.5){3}
  \Text(52,35)[l]{\bf Hot}
 \DashLine(150,17)(150,80){3}
  \Text(89,35)[l]{\bf Warm}
 \Text(123,35)[l]{\bf Cold}\LongArrow(150,55)(180,55)
 \Line(198.9,20)(198.9,130)
 \DashLine(175,27.5)(293,27.5){3}
 \Text(242,21.5)[l]{$T_F$}
\Text(200,40)[c]{$\Omega^{NTP}_{\tilde a}h^2\approx 1$}

 \Line(107.5,217.5)(234,85)
 \Curve{(234,85)(257.5,72)(300,60)}
 \DashLine(128,138)(153,163){3}\ArrowLine(149,159)(153,163)
 }

 \SetWidth{0.5}
 \DashLine(207.5,35)(207.5,110){3}\DashLine(207.5,35)(290,35){3}
 \Text(155,160)[l]{\color{yellow}\bf EXCLUDED}
\Text(155,145)[l]{\color{yellow}($\Omega^{TP}_{\tilde a}h^2> 1$)}
 \Text(130,102)[r]{($\Omega^{TP}_{\tilde a}h^2= 1$)}

 \end{picture}
\end{center}
 \caption{The solid line is the upper bound from TP. The yellow region is the region where NTP can give cosmologically interesting results ($\Omega_{\tilde a}^{\rm
NTP} h^2\simeq1$). The freezeout temperature is
$T_F\approx\frac{m_\chi}{20}$.}\label{fig:axinocdm}
\end{figure}

For axino to be the LSP, it must be lighter than the lightest neutralino and gravitino. Thus, the axino mass estimate is of prime importance. The conclusion is that there is no theoretical upper bound on the axino mass \cite{axinomass}, and we take it any value between 100 GeV and keV. Thus, there are two axino DM possibilities: warm DM with keV axinos \cite{Raja} and CDM with GeV axinos \cite{CKRosz}.
The gravitino problem is absent if gravitino is the next LSP (NLSP), $m_{\tilde a}<m_{3/2}<m_{\chi}$, since the thermally produced (TP) gravitinos would decay to axino and axion which would not affect the BBN produced light elements \cite{Yanag}. On the other hand, if the lightest neutralino is the NLSP, the TP mechanism restricts the reheating temperature after inflation. At the high reheating temperature, the TP is dominant in the axino production. If the reheating temperature is below the critical energy density line, there exists another axino CDM possibility by the nonthermally produced (NTP) axinos by the neutralino decay \cite{CKKR}. This situation is shown in Fig. \ref{fig:axinocdm}. Since the final axino energy fraction is reduced by the mass ratio, $\Omega_{\tilde a}h^2=(m_{\tilde a}/m_{\chi})\Omega_{\chi}h^2$ for $m_{\tilde a}<m_{\chi}<m_{3/2}$, the stringent cosmologically constrained MSSM parameter space for $m_\chi$ can be expanded.
%%%%%%%%%%%%%%%%%%%%%%%%%%%%%%%%%%%%%%%%%%%%%%%%%%%%%%%%%%%%%%%%%%%%%%%%%%%%
%%%%%%%%%%%%%%%%%%%%%%%%%%%%%%%%%%%%%%%%%%%%%%%%%%%%%%%%%%%%%%%%%%%%%%%%%%%%
%%%%%%%%%%%%%%%%%%%%%%%%%%%%%%%%%%%%%%%%%%%%%%%%%%%%%%%%%%%%%%%%%%%%%%%%%%%%%%
As shown in this figure, the NTP axinos can be CDM for relatively low reheating temperature ($<10$ TeV) for 10 MeV $<m_{\tilde a}<m_{\chi}$. The yellow region corresponds to the MSSM models with $\Omega_\chi h^2<10^4$, and a small axino mass renders the possibility of axino forming 23\% of the closure density. If all SUSY mass parameters are below 1 TeV, then probably $\Omega_\chi h^2<100$ (the grey region) but a sufficient axino energy density requires $m_{\tilde a}>1$ GeV. Thus, if the LHC does not detect the neutralino needed for its closing of the universe, the axino closing is a possibility \cite{KYChoi}. The efforts to detect axinos may be difficult \cite{axdetection}.

In the gauge mediated SUSY breaking (GMSB) scenario \cite{DineNelson}, the gravitino mass is generally smaller than the neutralino mass and possibly smaller than the axino mass, for which case cosmology has been studied \cite{AxinoGravitino}. Indeed, recently it has been shown that the GMSB is possible in the compactification of the heterotic string \cite{GMSBstring}.

\section{Axions from superstring}
As the final journey on axion, let us comment on its relevance in superstring theory. There are a few issues: the appearance of axions in superstring theory, a general difficulty of introducing a detectable QCD axion with an exact PQ symmetry, an approximate PQ symmetry, and search for the detectable QCD axion models along this line.

Superstring tells us a definite thing about global symmetries: there is no global symmetry. But the bosonic degrees from the
antisymmetric tensor field $B_{MN}$ may behave like pseudoscalars. If an axion is present, it is better to be one component in $B_{MN}$. Massless pseudoscalar degrees from $B_{MN}$ are classified as the model independent (MI) axion \cite{Witten84} and the model dependent (MD) axion \cite{Witten85}. The MI-axion is $B_{\mu\nu}\
(\mu,\nu=0,1,2,3)$  and MD axions are $B_{ij}\ (i,j=4,\cdots,9)$. It is known that MD axions are generally heavy \cite{WenW}, but it may be a model dependent statement.

The superstring axion decay constants are expected near the string scale which is too large compared to $10^{12}$ GeV. For the MI axion it has been calculated in \cite{ChoiKimF}, $F_a\sim 10^{16}$ GeV. So, a key question in superstring axion models toward the discovery {\it a la} the Sikivie type detectors is, $\lq\lq$How can one obtain a low value of $F_a$?"

An idea is the following. In some compactification schemes, an anomalous U(1) gauge symmetry results \cite{AnomU1}, where the U(1) gauge boson eats the MI axion so that the U(1) gauge boson becomes heavy. In fact, even before considering this anomalous U(1) gauge boson, the possibility was pointed out by Barr \cite{BarrAnom}, which became a consistent theory after discovering the anomalous U(1). Then, a global symmetry survives down the anomalous U(1) gauge boson scale. $F_a$, the breaking scale of this global symmetry, may be put in the axion window, which was stressed early in \cite{Kim88}, and recently by Svrcek and Witten \cite{Svrcek}. However, this idea on the decay constant does not work necessarily.

MD-axion decay constants were tried to be lowered by localizing them at fixed points \cite{Conlon,IWKim}. It uses the flux compactification idea and  is possible to have a small $F_a$ compared to the string scale by localizing the axion at a fixed point. Here, one needs a so-called Giddings-Kachru-Polchinski throat \cite{GKP}. However, even if one lowered some $F_a$, we must consider the hidden sector also in estimating the axion masses and decay constants. With the hidden confining force, we need two $\theta$s which have to be settled to zero and hence we need at least two axions. In this case, axion mixing must be considered. Here, there is an important (almost) theorem: the {\it cross theorem} on decay constants and condensation scales. Suppose that there are two axions $a_1$ with $F_1$ and $a_2$ with $F_2\ (F_1\ll F_2)$ which couple to  two nonabelian groups with scales $\Lambda_1$ and $\Lambda_2\ (\Lambda_1\ll \Lambda_2)$. The theorem states that \cite{kim99,IWKim}: according to the diagonalization process in most cases with general couplings, the larger condensation scale $\Lambda_2$ chooses the smaller decay constant $F_1$, and the smaller condensation scale $\Lambda_1$ chooses the larger decay constant $F_2$. So, just obtaining a small decay constant is not enough. The hidden sector may steal the smaller decay constant. It is likely that the QCD axion is left with the larger decay constant.

So, we must look for another possibility for a detectable QCD axion. This can be possible with an approximate PQ symmetry in string models. After all, the topologically attractive $B_{MN}$ may not be the QCD axion we want. There exists an earlier field theoretic work regarding an approximate PQ symmetry, starting with a discrete symmetry \cite{LShafi} where $Z_9$ was used. In string models, such approximate PQ symmetry was not calculated before. Since now we have an explicit model for the MSSM \cite{Kyae06,GMSBstring}, it is possible to consider an approximate symmetry. We can check whether this idea of approximate global symmetry is realized. For this idea to work, it is better that the PQ symmetry is preserved up to sufficiently higher orders. In this sense, $Z_{12}$ is helpful. In this vein, we calculated the $a\gamma\gamma$ coupling from the $Z_{12-I}$ superstring model \cite{Kyae06} for the first time \cite{ChoiIW}. This string calculation is shown as a line in Figs. \ref{CASTexp} and \ref{CosmicAxion}. Here, there are so many Yukawa couplings to be considered in a string derived model. For example, we encountered $O(10^4)$ terms for the $d=7$ superpotential and it is not a trivial task to find an approximate PQ symmetry direction. Noting that the MI axion with anomalous U(1) always has a large decay constant since all fields are charged under this anomalous U(1), a phenomenologically observable QCD axion must need an approximate PQ symmetry.

\section{Conclusion}

The popular CDM candidates are WIMPs and very light axions. We know that they must fill the universe based on the observational grounds and by the fact that we exist here in a planet. Direct searches for WIMPs in the universe use the WIMP cross section in our environment. The LHC machine will tell whether the LSP mass falls in the CDM needed range or not. The other candidate a very light axion, whether or not it is the dominant CDM component, is believed to exist from the need for a solution of the strong CP problem. So, I reviewed the axion and the related issues: solutions of the strong CP problem, the axion CDM possibility, the axino CDM possibility with  order GeV axino mass, and an observable QCD axion possibility from superstring.

Maybe solar axions are easier to be detected than cosmic axions. Then, axion is not the dominant component of CDM.  Most exciting however would be that axion is discovered and its discovery confirms instanton physics of QCD (by experiments).\\

\noindent {\bf Acknowledgments\ :} This work is supported in part by the KRF Grants, No. R14-2003-012-01001-0 and No. KRF-2005-084-C00001.


\begin{thebibliography}{99}
\bibitem{verylight} {Once it was called an invisible axion but is better to be called a very light axion due to the possibility of its detection.}
\end{thebibliography}
\end{document}